\documentclass[aps,prd,12point,twocolumn,nofootinbib,showpacs,superscriptaddress]{revtex4-2} 
\def\theequation{\arabic{section}.\arabic{equation}} \usepackage{psfrag} 
\usepackage{subfigure} \usepackage{color} \usepackage{mathrsfs} 
\usepackage{graphicx} \usepackage{amssymb, bm} \usepackage{amsmath, 
amsthm} \usepackage{epstopdf} \usepackage{hyperref} \usepackage{enumerate} 
\usepackage{longtable} \usepackage{amsmath} \usepackage[normalem]{ulem} 
\usepackage[title]{appendix}

\newcommand{\be}{\begin{equation}} \newcommand{\ee}{\end{equation}}

\definecolor{pinegreen}{rgb}{0.0, 0.47, 0.44}

\begin{document} \def\theequation{\arabic{section}.\arabic{equation}}


\title{Black hole event horizons are cosmologically coupled}

\author{Valerio Faraoni} \email[]{vfaraoni@ubishops.ca}
\affiliation{Department of Physics \& Astronomy, Bishop's University, 2600 
College Street, Sherbrooke, Qu\'ebec, Canada J1M~1Z7}

\author{Massimiliano Rinaldi} \email[]{massimiliano.rinaldi@unitn.it}
\affiliation{Department of Physics, University of Trento, Via 
Sommarive 14, 38122 Trento, Italy} 
\affiliation{Trento Institute for Fundamental 
Physics and Applications TIFPA-INFN, Via 
Sommarive 14, 38122 Trento, Italy}

\begin{abstract}

It is shown that an exactly static and spherically symmetric black 
hole event horizon cannot be embedded in a time-dependent geometry. 
Forcing it to do so results in a naked null singularity at the would-be 
horizon. Therefore, since the universe is expanding, black holes must 
couple to the cosmological expansion, which was suggested as the growth 
mechanism for supermassive black holes in galaxies, with implications for 
the dark energy puzzle.

\end{abstract}

\maketitle

\section{Introduction} 
\label{sec:1} 
\setcounter{equation}{0}

A major problem of contemporary astrophysics consists of the fact that 
supermassive black holes at the centres of galaxies must have grown very 
fast { at redshifts $z>6$}, which is problematic if one invokes 
standard astrophysical channels 
\cite{Inayoshi:2019fun, Volonteri:2021sfo}. Moreover, LIGO detections of 
gravitational waves from binary mergers involving black holes 
highlight the presence of stellar mass black holes in the mass 
gap where they should not exist according to well-established models of stellar 
evolution \cite{LIGOScientific:2020iuh, KAGRA:2021vkt, Mehta:2021fgz}{, which is another potential problem if confirmed.}   
These astrophysical puzzles can potentially be resolved if black holes 
interact with the universe in which they are embedded and their masses 
evolve over cosmological time scales (``cosmological 
coupling'') \cite{Farrah:2023opk} \footnote{{ To be specific, in this paper  
``cosmological coupling'' refers to {\em any} evolution of a black hole 
event horizon with time or with the scale factor of the FLRW in which they 
are embedded.}  }.

{ These astrophysical black holes are usually modelled as 
asymptotically flat Schwarzschild or Kerr black holes. However, 
r}ealistic black holes are not asymptotically flat but are embedded in the 
universe. The study of astrophysical black holes usually neglects the 
cosmic expansion due to the vast separation between astrophysical (local) 
scales and the cosmological scale. However, over billions of years, tiny 
interactions between black hole and surrounding universe, if present, 
accumulate and can become important. The recent Ref.~\cite{Farrah:2023opk} 
reports  plausible evidence for
cosmological coupling between the masses of 
supermassive black holes and the scale factor of the 
Friedmann-Lema\^itre-Robertson-Walker (FLRW) universe, obtained from data 
of black hole populations in red elliptical galaxies in the redshift range 
$0<z\leq 2.5$. Furthermore, if black holes are non-singular objects with 
de Sitter-like macroscopic cores, they could even explain dark energy, 
which would effectively be relegated to the interiors of black holes 
\cite{Farrah:2023opk}, a phenomenology possibly extending to non-singular 
black hole mimickers, which are now the subject of a vast theoretical 
literature  (e.g., 
\cite{Bardeen68,Dymnikova:1992ux,Hayward:2005gi, Mazur:2001fv, 
Ansoldi:2008jw, Franzin:2023slm, Cadoni:2023lqe,Cadoni:2023lum, 
Cadoni:2024jxy}.  Cosmological 
coupling is also in agreement with the recent DESI results on the redshift 
evolution of the dark energy equation of state \cite{Croker:2024jfg}. The 
2023 tentative observation of cosmological coupling \cite{Farrah:2023opk} 
caused a lively debate in both the observational and theoretical 
communities during the past year { 
(\cite{Rodriguez:2023gaa,Andrae:2023wge,Gao:2023keg,Amendola:2023ays, 
Lacy:2023kbb} and references therein)}, as the arguments supporting it 
seemed unconvincing to some of its critics - here we provide more convincing 
ones.

Another context in which black holes can potentially 
interact with their cosmological surroundings is given by  
primordial black holes in the early universe, in which the Hubble scale is 
small and the separation between black hole and cosmological scales may 
not be sharp. Ref.~\cite{Boehm:2020jwd} pointed out the inadequacy of the 
Schwarzschild (or Kerr) metric to describe primordial black holes. 
Although the analysis relied on the specific Thakurta solution of the 
Einstein equations that ultimately turns out not to describe a black hole 
\cite{Harada:2021xze,Hutsi:2021nvs}, 
its main point remains.

The problem of principle of describing a black hole or other strong field 
object immersed in a FLRW universe has a long history dating back to 
McVittie \cite{McVittie:1933zz} and Einstein and Straus 
\cite{Einstein:1945id} (see \cite{Faraoni:2015ula} for a review of 
relevant solutions of the Einstein equations). Only recently, however, 
cosmological coupling (independent of accretion or mergers)  has been 
included in realistic astrophysics (\cite{Farrah:2023opk,Croker:2021duf} 
and references therein). In spite of the extremely interesting potential 
implications, the cosmological coupling of black holes does not yet have 
solid theoretical foundations, at least locally, nor it is backed by exact 
solutions of the Einstein equations possessing all the desirable 
properties for coupling, although progress was made in this direction 
\cite{Croker:2019mup,Croker:2019kje}.  The possibility that black holes 
couple to the cosmological expansion, and that this coupling has even been 
observed, has triggered a debate in the community. Critics point to exact 
coupled solutions without any mass growth such as the Schwarzschild-de 
Sitter/Kottler solution (e.g., \cite{Gaur:2023hmk,Dahal:2023suw}). A 
counter-argument is that, being general relativity a nonlinear theory, the 
solutions of its initial/boundary value problems are not always unique 
(e.g., \cite{Croker:2024jfg}). To address this issue, here we turn around 
the question and ask whether an exactly static black hole event horizon (a 
null surface and a causal boundary) can be embedded in a FLRW 
universe.\footnote{Because it is locally static, we exclude the de Sitter 
universe from our discussion (indeed, the Schwarzschild-de Sitter/Kottler 
solution is locally static and contains a black hole event horizon, but it 
is irrelevant for the problem addressed here).} Interestingly, this 
question was raised in \cite{Davidson:2012si}, when no one was yet taking 
cosmological coupling seriously. These authors assume a solution of the 
Einstein equations that is spherically symmetric, time-dependent, and is 
sourced by a perfect fluid with linear barotropic equation of state $P=w 
\rho$ (with $w=$~const.), answering the question negatively through a 
perturbative analysis \cite{Davidson:2012si}.  Here we generalise 
significantly the result of \cite{Davidson:2012si} by studying the 
existence of an exactly static black hole horizon in a generic 
time-dependent geometry and provide several no-go arguments.  In 
particular, we do not assume the Einstein equations nor specific forms of 
matter, but we rely only on a near-horizon expansion well-established in 
black hole physics. We find that embedding an exactly static black hole 
horizon in a time-dependent geometry turns it into a naked null 
singularity, providing theoretical support for cosmological coupling.

The next section sets up the description of an exactly static black 
hole event horizon in a time-dependent geometry, while the following 
sections show how this assumption leads to absurdities. Specifically, 
Sec.~\ref{sec:3}, shows that null and timelike radial geodesics are 
incomplete at the would-be static horizon, where the Ricci 
scalar also diverges. This static horizon must coincide 
with a naked spacetime singularity at a finite (proper) radius. In 
Sec.~\ref{sec:4} we analyze the propagation of a test scalar field, which 
exhibits divergences at the would-be static horizon. Sec.~\ref{sec:5} 
assumes a FLRW ``background'' and argues that the static would-be horizon 
leads to a paradox for the Hawking temperature, while  Sec.~\ref{sec:6} 
contains the conclusions.

\section{Static black hole event horizon in a time-dependent metric} 
\label{sec:2} 
\setcounter{equation}{0}

Let us set up the description of a static black hole event 
horizon embedded in a time-dependent ``background'' 
geometry which, remembering the origin of our problem, can be thought of 
as a FLRW universe. However, we 
do not make this assumption until Sec.~\ref{sec:5} and the time-dependent 
metric is  
general, except for being spherically symmetric.
We do not want to commit to any exact solution of the Einstein equations 
(or of the field equations of other theories of gravity) because these are 
{\em ad hoc}, and only a handful of them are known anyway 
\cite{Faraoni:2015ula}. In fact, we do not even assume the Einstein 
equations.

Without loss of generality, a spherically symmetric and 
time-dependent line element  can be written in the form
\be
ds^2=-T^2(t,r) dt^2 +a^2(t,r) \left( dr^2 +r^2 d\Omega_{(2)}^2 \right) 
\label{metric}
\ee
in isotropic coordinates. $ d\Omega_{(2)}^2 \equiv d\vartheta^2 + 
\sin^2 
\vartheta d\varphi^2$ is the line element on the unit 2-sphere and the 
areal radius (a geometric quantity defined invariantly) is 
\be 
R(t,r) = a(t,r) r \,,\label{AHlocator}
\ee 
as can be read off the angular part of the line element~(\ref{metric}). 
The equation locating the apparent horizons\footnote{This equation 
generalizes the well known relation locating the Schwarzschild event 
horizon $g^{RR}=\nabla^cR \nabla_cR=1-2m/R=0$ to any spherical geometry.
{ In spherical symmetry, the condition $\nabla^c R \nabla_c R=0$ is 
equivalent to  
the standard definition of apparent horizon in terms of the  
expansion scalars of ingoing and outgoing null geodesic congruences 
\cite{Abreu:2010ru,Faraoni:2015ula}.}} 
(if they exist) is 
$\nabla^c 
R \nabla_c R=0$. A single root corresponds to a black hole or white hole 
horizon, a double root to a wormhole \cite{Faraoni:2015ula}. This 
equation locates also event horizons, and  gives
\be
g^{tt} \dot{R}^2 + g^{rr} R'^2 =  
-\frac{\dot{R}^2}{T^2}+ \frac{R'^2}{a^2}=0 \label{questa}
\ee
where $\dot{R} \equiv \partial R/\partial t$, $ R' \equiv \partial 
R/\partial r$. Equation~(\ref{questa}) is obtained also by imposing that a 
spherical surface of 
radius $R$ be null: $N_a\equiv \nabla_a R $ with $N^a N_a=0$. Indeed, it 
is well known that 
static apparent horizons coincide with null event horizons.


{ At a fixed instant of time $t=$~const., the event horizon is a 
2-dimensional surface, a 2-sphere $R=$~const. spanned by the angular 
coordinates $\left( \vartheta, \varphi \right)$.} 
The two-dimensional metric induced on the surface $R=$~const. is static 
but the time-dependent geometry just outside the horizon has $T=T(t,r)$, 
$a=a(t,r)$ and the areal radius $R(t,r)=a(t, r) r$ varies in time and 
cannot match $R=$~const. on the would-be horizon. Therefore, the junction 
conditions \cite{Barrabes:1991ng, Poisson:2002nv} cannot be satisfied 
there: the discontinuity in the metric $g_{ab}$ induced on this 
2-surface cannot be handled with 
distributional techniques and the jump in the metric $g_{ab}$, unlike a 
jump in its transversal derivative (interpreted as due to a matter layer 
on a null  shell) has no physical interpretation \cite{Poisson:2002nv}.

\section{Radial geodesics and Ricci scalar at the horizon} 
\label{sec:3} 
\setcounter{equation}{0}

Let us examine outgoing and ingoing radial null { and timelike} geodesics. We will show 
that they cannot be continued inside the would-be static black hole 
horizon, which is unphysical. Furthermore, the Ricci scalar of the 
geometry~(\ref{metric}) diverges there. The difficulties arise from 
assuming an exactly static horizon in a time-dependent geometry, showing 
that this assumption is unphysical.

\subsection{Radial null geodesics}

Consider radial null geodesics parametrized by an affine parameter 
$\lambda$ and with 4-tangent $u^a= \left( u^0 , u^1, 
0, 0 \right)$. The normalization $u_c u^c=0$ reads
\be
- T^2 ( u^0)^2 + a^2 (u^1)^2 =0 \,,
\ee
yielding the coordinate velocity of light rays 
\be
\frac{u^1}{u^0}=\frac{dr}{dt} = \pm \frac{T}{a} 
\,,\label{coordinatevelocity}
\ee
where the upper sign refers to outgoing, and the lower one to ingoing, 
radial null geodesics. The equation of these affinely parameterized 
geodesics is 
\be
\frac{d u^a}{d\lambda} + \Gamma^a_{bc} u^b u^c =0 
\,. \label{radialnullgeodesicseq} 
\ee
Using~(\ref{coordinatevelocity}), the time component 
gives 
\be
\frac{du^0}{d\lambda}+ \left( \Gamma^0_{00} \pm \frac{2T}{a} \, 
\Gamma^0_{01} + \frac{T^2}{a^2} \, \Gamma^0_{11} \right) ( u^0)^2=0 
\ee 
which, using the Christoffel symbols of the metric~(\ref{metric})
\be
\Gamma^0_{00} = \frac{\dot{T}}{T} \,, \quad 
\Gamma^0_{01}  = \frac{T'}{T} \,, \quad
\Gamma^0_{11} = \frac{a\dot{a}}{T^2} \,,
\ee
assumes the form
\be
\frac{d}{d\lambda} \left( \frac{1}{u^0} \right) =
\frac{\dot{T}}{T}  \pm \frac{2T'}{a} 
+ \frac{ \dot{a}}{a} \,.\label{cazzo}
\ee
Similarly, using the Christoffel symbols
\be
\Gamma^1_{00} = \frac{TT'}{a^2} \,, \quad
\Gamma^1_{01}  = \frac{ \dot{a}}{a} \,, \quad 
\Gamma^1_{11} = \frac{a'}{a} \,,
\ee
the radial component of the geodesic 
equation~(\ref{radialnullgeodesicseq})  gives
\be
\frac{d}{d\lambda} \left( \frac{1}{u^1}\right) =  \frac{T'}{T}  \pm 
\frac{2\dot{a}}{T} 
+ \frac{a'}{a}  \,. \label{u1null}
\ee 
Close to the static event horizon, the spherical geometry must resemble 
Schwarzschild which, in isotropic 
coordinates, is well known to be approximated by the Rindler 
metric 
\be
ds^2 = ds^2_\mathrm{Rindler} + \, ... = - \frac{x^2}{4} dt^2 + 4m^2 \left( 
dx^2 + d\Omega_{(2)}^2 \right) + \, ... \,,
\ee
where $m$ is the black hole mass and $x$, given by $ r=\frac{m}{2} 
\left(1+x \right)$ (with $0<x \ll 1$), measures the radial 
distance to the black hole horizon  
\cite{FabbriNavarro-Salas}. The horizon  geometry  
is {\em exactly} static, hence $\dot{T}=\dot{a}=0$ there and 
Eq.~(\ref{cazzo}) 
approximates to
\be
\frac{d}{d\lambda} \left( \frac{1}{u^0}\right) \simeq \pm \frac{2T'}{a} 
\simeq \pm \frac{2}{a_\mathrm{H}} 
\ee
since $T'\simeq 1$, which yields $
u^0 \simeq \frac{C}{\lambda -\lambda_0} $, 
with $C$ and $\lambda_0$ constants chosen so that $u^0$ is positive to 
keep the 4-tangent future-oriented. 
Then, either  $u^0 
\to 0 $ as $\lambda \to +\infty$, i.e., ``time stops'' (if 
$\lambda>\lambda_0$ and $C>0$) or else the radial null 
geodesic stops at 
the finite value $\lambda_0$ of $\lambda$ (if 
$\lambda < \lambda_0$ and $C<0$), with $u^0$ still diverging. On the 
contrary, $u^1= \pm 
Tu^0 / a \to 0$ and 
the coordinate velocity~(\ref{coordinatevelocity}) tends to zero 
as $  x\to 0^{+}$, 
i.e., the radial motion of light stops at the horizon (for ingoing radial 
null geodesics) or cannot begin there (for outgoing ones). The fact that  
ingoing radial photons  
cannot penetrate the static event horizon and fall into the black hole is 
unphysical. {The study of radial timelike geodesics gives similar 
results. }

{
\subsection{Radial timelike geodesics}

We now turn to timelike geodesics. The 4-velocity $u^a=dx^a/d\tau$ (where $\tau$ is the proper 
time) tangent to a radial timelike geodesic has  
components $\left( u^0, u^1, 0, 0 \right)$ and is normalized to $u^c 
u_c=-1$, 
which gives 
\be
T^2 \left(u^0\right)^2 =a^2 \left(u^1 \right)^2 +1 
\,. \label{normalization2}
\ee
Since $(u^0)^2 =\frac{ 1+ (a u^1)^2}{T^2} \geq \frac{1}{T^2} \to +\infty$ 
as 
$x\to0$, $u^0$ diverges as least as fast as $T^{-1}=2/x$ at the static 
horizon. The time and radial components of the geodesic equation are 
(using $\dot{a}, T\to 0$)
\begin{eqnarray}
&& \frac{ du^0}{d\tau} \pm \frac{2u^0}{a} \, \sqrt{ T^2 (u^0)^2 -1} \, 
\frac{T'}{T} = 0 \,, \\
&&\nonumber\\
&& \frac{ d u^1}{d\tau} \pm \frac{1+ a^2 (u^1)^2}{a^2} \,  \frac{T'}{T} = 
\frac{a'}{a^3}  \label{timelikeu1}\,,
\end{eqnarray}
where, in~(\ref{timelikeu1}), we used the 
normalization~(\ref{normalization2})  $u^0= \sqrt{ 1+\left( a 
u^1\right)^2}/T$. Near the horizon, Eq.~(\ref{timelikeu1}) becomes
\be
\frac{du^1/d\tau}{ (u^1)^2 +1/a^2} = -\frac{1}{T} \,,
\ee
which can be integrated at the horizon where $T$ is static, obtaining
\be
u^1 \simeq \frac{1}{a} \tan \left( \frac{\tau_0-  \tau}{aT}\right) \,, 
\ee
where $\tau_0$ is an integration constant. As $x\to 0 $, $u^1$ is only 
defined for  a finite range of the proper time $\tau$. Radially infalling 
objects stop at the horizon and cannot enter it.
}

\subsection{Ricci scalar}


Geodesic incompleteness is the 
signature of spacetime singularities in the Hawking-Penrose singularity 
theorems \cite{Hawking:1973uf, Wald:1984rg}, signaling the 
breakdown 
of spacetime at the static horizon. The other common definition of 
spacetime 
singularity, the divergence of curvature scalars, confirms this 
conclusion.

The Ricci scalar of the static geometry~(\ref{metric}) (not to be confused 
with the areal radius $R$)  is
\begin{eqnarray}
{\cal R} &=& -\frac{4a''}{a^3} +\frac{6\ddot{a}}{aT^2} -\frac{2a'T'}{a^3T} 
-\frac{6\dot{a}\dot{T}}{aT^3} +\frac{6 \dot{a}^2}{a^2 T^2} -\frac{8a'}{a^3 
r} \nonumber\\
&&\nonumber\\
&\, & -\frac{4T'}{a^2 r T} 
- \frac{2T''}{a^2 T} +\frac{2a'^2}{a^4} \,. \label{pre-Ricci}
\end{eqnarray}
Because the horizon is exactly static, it must be $\dot{a}=0$ and 
$\dot{T}=0$ in the limit $x\to 0$, where~(\ref{pre-Ricci}) 
reduces to 
\be
{\cal R} \simeq  -\frac{4a''}{a^3}  
-\frac{2a'T'}{a^3T}  -\frac{8a'}{a^3 
r} -\frac{4T'}{a^2 r T} - \frac{2T''}{a^2 T} +\frac{2a'^2}{a^4} \,. 
\label{cazzo1}
\ee
The first and the last terms in the right-hand side of Eq.~(\ref{cazzo1}) 
are finite and can be neglected in comparison with the terms diverging as 
$1/T$; using $T \sim x/2$, $T'\simeq 1$, $T''\simeq 0$ and denoting with 
$r_\mathrm{H}$ and $a_\mathrm{H}$ the 
finite values of $r$ and $a$ on the horizon, one obtains
\be
{\cal R} \simeq -\frac{2}{a_\mathrm{H}^2} \left[ \left( 
\frac{a_\mathrm{H}'}{a_\mathrm{H}} 
+\frac{2}{r_\mathrm{H}}\right) \frac{T'}{T} +\frac{T''}{a_\mathrm{H} T} 
\right] 
\simeq 
\frac{1}{x} \to \infty 
\ee
as $ x\to 0^{+} $. Similar considerations hold for the Kretschmann 
and other curvature scalars.

\section{Test Klein-Gordon field} 
\label{sec:4} 
\setcounter{equation}{0}

Matter is also subject to divergences at the would-be static 
horizon. Let us consider  a free massless test scalar field $\phi$ 
satisfying the Klein-Gordon equation
\be
\Box \phi = \frac{1}{\sqrt{-g}}\, \partial_a  \left( \sqrt{-g} \, g^{ab} 
\partial_b \phi \right)=0 \,,\label{KleinGordon}
\ee
where $g=\mbox{Det}(g_{ab})$.  
We are interested in the near-horizon propagation  of the $\phi$-waves  
and, for simplicity, we restrict to $s$-modes with  
$\phi=\phi(t,r)$. In the geometry~(\ref{metric}), 
Eq.~(\ref{KleinGordon}) reads
\be
\ddot{\phi} +\left( \frac{3\dot{a}}{a} +\frac{ \dot{T}}{T} \right) 
\dot{\phi} 
-\frac{T^2}{a^2} \, \phi'' -\frac{T}{a^2} \left( T' +\frac{ 
Ta'}{a} 
+\frac{2T}{r} \right) \phi' =0 \,.
\ee
Since the would-be horizon is exactly static, $\dot{a}=0$, $ 
\dot{T}=0$ and $T\to 0$, at the horizon the 
$s$-waves satisfy $\ddot{\phi} \simeq 0$ and $\phi(t)=\phi_0+\phi_1 t $ 
(where $\phi_{0,1}$ are constants), diverging as  
$t\to+\infty$. 
Their energy-momentum tensor $ 
 T_{ab}=\nabla_a \phi 
\nabla_b \phi -\frac{1}{2} \, g_{ab} \nabla^c \phi \nabla_c \phi$ has 
trace
\be
{T^a}_a= \left( \frac{\dot{\phi}}{T} \right)^2 \to +\infty
 \quad \quad \mbox{as } \; x\to 0^{+} \,.
\ee
The free scalar field is equivalent to an effective perfect fluid with  
stiff equation of state, effective energy density 
$\rho $, effective 
pressure $P$, and trace $T=-\rho+3P$, which 
diverges at the static 
horizon. Since $\rho= T_{00} = \dot{\phi}^2/2 $ is finite, the effective 
pressure diverges at the horizon.

\section{Hawking temperature}
\label{sec:5} 
\setcounter{equation}{0}

Assume now that the time-dependent geometry is 
asymptotically FLRW, describing an exactly static black hole horizon 
embedded in a FLRW universe.  The Hawking temperature at the horizon is 
$T_\mathrm{H}=1/(8\pi 
m)$, 
where 
$m$ is the black hole mass,  taken as the 
Misner-Sharp-Hernandez mass defined in spherical symmetry (which reduces 
to the Schwarzschild mass) and must be constant for the black hole 
horizon to remain static. The black hole must be in 
equilibrium with its Hawking radiation, but the temperature of a 
photon gas in local thermal equilibrium when decoupled from other cosmic 
fluids scales as 
$T\sim 1/a$ to maintain the blackbody distribution. To wit, a physical 
wavelength scales as $\lambda_\mathrm{phys}=\lambda_\mathrm{c} a$, where 
$\lambda_\mathrm{c}$ is 
the comoving wavelength, the frequency $\nu=c/\lambda_\mathrm{phys} \sim 
1/a$ 
and the Planck distribution of the spectral energy density
\be
u(\nu, T) =\frac{8\pi h \nu^3}{c^3} \, \frac{1}{\mbox{e}^{ 
\frac{h\nu}{K_BT} 
}-1} 
\ee
(where $h$ and $K_B$ are the Planck and the Boltzmann constants, 
respectively) is only preserved if  $T\sim 1/a$. 

Furthermore, the analysis of exact solutions of the Einstein equations 
representing black holes embedded in FLRW universes, which are usually 
described by dynamical {\em apparent} (but not event) horizons, yields 
black hole temperatures scaling as $T \sim \frac{1}{8\pi ma}$. This is the 
case for cosmological black hole solutions of the Einstein equations 
obtained by conformally transforming Schwarzschild \cite{Faraoni:2007gq}, 
in particular for the Sultana-Dyer black hole scrutinized 
in greater detail \cite{Saida:2007ru}. At our would-be static horizon it  
must be simultaneously $T=1/(8\pi m )=$~const. and $T\sim 1/a(t)$, which 
is impossible. 


\section{Conclusions} 
\label{sec:6} 
\setcounter{equation}{0}

An exactly static black hole event horizon embedded in the time-dependent 
geometry~(\ref{metric}) is, in reality, a naked null spacetime singularity 
located at a finite areal radius.

All radial null and timelike geodesics are 
future-inextendible at this static horizon.\footnote{The Hawking-Penrose  
theorems identify a spacetime singularity where {\em at least  
one} geodesic is future-inextendible \cite{Hawking:1973uf, Wald:1984rg}.} 
Geodesic incompleteness defines spacetime singularities in 
the 
Hawking-Penrose singularity theorems \cite{Hawking:1973uf, Wald:1984rg}. 
 The Ricci scalar also diverges at the would-be horizon.  A radially 
propagating test scalar field has divergent effective pressure there. 
Finally, the Hawking temperature of a strictly static black hole horizon 
would have to be constant, while the expansion of the universe in which it 
is embedded dictates that it must scale like the inverse scale factor, 
which are incompatible requirements.

The analysis in \cite{Davidson:2012si} assumed the Einstein equations and 
a perfect fluid with linear barotropic equation of state $P=\mbox{const.} 
\times \rho$ as a source. Here, we significantly generalized that result 
and we did not use any of these assumptions. Furthermore, we did not 
assume that the geometry is asymptotically FLRW (except in Sec. V) but 
only that it is time-dependent outside the static event horizon, therefore 
our result is quite general, including
 FLRW as a special case.

Forcing an exactly static black hole event horizon onto a dynamic 
``background'' (e.g., a FLRW universe) turns this horizon into a naked 
null singularity. Preventing it from expanding (which can only  be done 
with strong gravity objects since weak-field objects will comove) tears 
spacetime. The problems described and the unavoidability of the naked null 
singularity arise from the assumption that the black hole horizon is 
exactly static and disappear if this assumption is dropped, thus 
time-dependent event horizons are possible. Finding their dependence on 
time or on the scale factor of the host FLRW universe presumably requires 
the complete specification of the geometry and committment to specific 
metrics, and will be investigated in the near future.

{ The prospects of embedding an exactly stationary black hole event 
horizon in a FLRW (or any time-dependent) ``background'' are even more 
dire than for a spherical static horizon. In fact, a FLRW universe is not 
stationary and matching a horizon with axisymmetry (or any symmetry) with 
another metric that does not share this symmetry is doomed to failure. 
This fact is well known when trying to generalize the Swiss-cheese model 
(a  Schwarzschild region matched to spherically symmetric FLRW on a 
2-sphere) to axisymmetry. Matching a Kerr region to FLRW fails  
because the interior and exterior regions do not share axisymmetry  
\cite{Senovilla:1997zz,Mars:1998sq,Mars:2013ooa}. This consideration, 
which will be discussed further in the future, extends the physical 
relevance of the result presented here. 

Finally, we have not determined the precise way in which black hole event 
horizons embedded in given FLRW universes evolve with time. This law would 
be most interesting for astrophysics to decide whether existing puzzles 
can be solved by cosmological coupling, and will be addressed in the 
future.}

\begin{acknowledgments} We are grateful to Kevin Croker { and Duncan 
Farrah} for discussions.  V.F. is supported by the Natural Sciences \& 
Engineering Research Council of Canada (grant 2023-03234).

\end{acknowledgments}






\begin{thebibliography}{104}

\bibitem{Inayoshi:2019fun}
K.~Inayoshi, E.~Visbal and Z.~Haiman,
``The Assembly of the First Massive Black Holes,''
Ann. Rev. Astron. Astrophys. \textbf{58}, 27-97 (2020)
doi:10.1146/annurev-astro-120419-014455
[arXiv:1911.05791 [astro-ph.GA]].

\bibitem{Volonteri:2021sfo}
M.~Volonteri, M.~Habouzit and M.~Colpi,
``The origins of massive black holes,''
Nature Rev. Phys. \textbf{3}, no.11, 732-743 (2021)
doi:10.1038/s42254-021-00364-9
[arXiv:2110.10175 [astro-ph.GA]].

\bibitem{LIGOScientific:2020iuh}
R.~Abbott \textit{et al.} [LIGO Scientific and Virgo],
``GW190521: A Binary Black Hole Merger with a Total Mass of $150  
M_{\odot}$,''
Phys. Rev. Lett. \textbf{125}, no.10, 101102 (2020)
doi:10.1103/PhysRevLett.125.101102
[arXiv:2009.01075 [gr-qc]].

\bibitem{KAGRA:2021vkt}
R.~Abbott \textit{et al.} [KAGRA, VIRGO and LIGO Scientific],
``GWTC-3: Compact Binary Coalescences Observed by LIGO and Virgo during 
the Second Part of the Third Observing Run,''
Phys. Rev. X \textbf{13}, no.4, 041039 (2023)
doi:10.1103/PhysRevX.13.041039
[arXiv:2111.03606 [gr-qc]].

\bibitem{Mehta:2021fgz}
A.~K.~Mehta, A.~Buonanno, J.~Gair, M.~C.~Miller, E.~Farag, R.~J.~deBoer, 
M.~Wiescher and F.~X.~Timmes,
``Observing Intermediate-mass Black Holes and the Upper Stellar-mass gap 
with LIGO and Virgo,''
Astrophys. J. \textbf{924}, no.1, 39 (2022)
doi:10.3847/1538-4357/ac3130
[arXiv:2105.06366 [gr-qc]].

\bibitem{Farrah:2023opk}
D.~Farrah, K.~S.~Croker, G.~Tarl\'e, V.~Faraoni, S.~Petty, J.~Afonso, 
N.~Fernandez, K.~A.~Nishimura, C.~Pearson and L.~Wang, \textit{et al.}
``Observational Evidence for Cosmological Coupling of Black Holes and its 
Implications for an Astrophysical Source of Dark Energy,''
Astrophys. J. Lett. \textbf{944}, no.2, L31 (2023)
doi:10.3847/2041-8213/acb704
[arXiv:2302.07878 [astro-ph.CO]].

\bibitem{Bardeen68} J. M. Bardeen, ``Non-singular general-relativistic 
gravitational collapse'', in {\em Proceedings of the International 
Conference GR5}, Tbilisi, USSR 1968 
(Tbilisi University Press, 1968).

\bibitem{Dymnikova:1992ux}
I.~Dymnikova,
``Vacuum nonsingular black hole,''
Gen. Rel. Grav. \textbf{24}, 235-242 (1992)
doi:10.1007/BF00760226

\bibitem{Hayward:2005gi}
S.~A.~Hayward,
``Formation and evaporation of regular black holes,''
Phys. Rev. Lett. \textbf{96}, 031103 (2006)
doi:10.1103/PhysRevLett.96.031103
[arXiv:gr-qc/0506126 [gr-qc]].

\bibitem{Mazur:2001fv}
P.~O.~Mazur and E.~Mottola,
``Gravitational Condensate Stars: An Alternative to Black Holes,''
Universe \textbf{9}, no.2, 88 (2023)
doi:10.3390/universe9020088
[arXiv:gr-qc/0109035 [gr-qc]].

\bibitem{Ansoldi:2008jw}
S.~Ansoldi,
``Spherical black holes with regular center: A Review of existing models 
including a recent realization with Gaussian sources,''
[arXiv:0802.0330 [gr-qc]].

\bibitem{Carballo-Rubio:2023mvr}
R.~Carballo-Rubio, F.~Di Filippo, S.~Liberati and M.~Visser,
``Singularity-free gravitational collapse: From regular black holes to 
horizonless objects,''
[arXiv:2302.00028 [gr-qc]].

\bibitem{Franzin:2023slm}
E.~Franzin, S.~Liberati and V.~Vellucci,
``From regular black holes to horizonless objects: quasi-normal modes, 
instabilities and spectroscopy,''
JCAP \textbf{01}, 020 (2024)
doi:10.1088/1475-7516/2024/01/020
[arXiv:2310.11990 [gr-qc]].

\bibitem{Cadoni:2023lqe}
M.~Cadoni, R.~Murgia, M.~Pitzalis and A.~P.~Sanna,
``Quasi-local masses and cosmological coupling of black holes and 
mimickers,''
JCAP \textbf{03}, 026 (2024)
doi:10.1088/1475-7516/2024/03/026
[arXiv:2309.16444 [gr-qc]].

\bibitem{Cadoni:2023lum}
M.~Cadoni, A.~P.~Sanna, M.~Pitzalis, B.~Banerjee, R.~Murgia, N.~Hazra and 
M.~Branchesi,
``Cosmological coupling of nonsingular black holes,''
JCAP \textbf{11}, 007 (2023)
doi:10.1088/1475-7516/2023/11/007
[arXiv:2306.11588 [gr-qc]].

\bibitem{Cadoni:2024jxy}
M.~Cadoni, M.~Pitzalis, D.~C.~Rodrigues and A.~P.~Sanna,
``Cosmological coupling of local gravitational systems,''
[arXiv:2406.06091 [gr-qc]].

\bibitem{Croker:2024jfg}
K.~S.~Croker, G.~Tarl\'e, S.~P.~Ahlen, B.~G.~Cartwright, D.~Farrah, 
N.~Fernandez and R.~A.~Windhorst,
``DESI Dark Energy Time Evolution is Recovered by Cosmologically Coupled 
Black Holes,''
[arXiv:2405.12282 [astro-ph.CO]].

{ 
\bibitem{Rodriguez:2023gaa}
C.~L.~Rodriguez,
``Constraints on the Cosmological Coupling of Black Holes from the 
Globular Cluster NGC 3201,''
Astrophys. J. Lett. \textbf{947}, no.1, L12 (2023)
doi:10.3847/2041-8213/acc9b6
[arXiv:2302.12386 [astro-ph.CO]].

\bibitem{Andrae:2023wge}
R.~Andrae and K.~El-Badry,
``Constraints on the cosmological coupling of black holes from Gaia,''
Astron. Astrophys. \textbf{673}, L10 (2023)
doi:10.1051/0004-6361/202346350
[arXiv:2305.01307 [astro-ph.CO]].

\bibitem{Gao:2023keg}
S.~J.~Gao and X.~D.~Li,
``Can Cosmologically Coupled Mass Growth of Black Holes Solve the Mass 
Gap Problem?,''
Astrophys. J. \textbf{956}, no.2, 128 (2023)
doi:10.3847/1538-4357/ace890
[arXiv:2307.10708 [astro-ph.HE]].

\bibitem{Amendola:2023ays}
L.~Amendola, D.~C.~Rodrigues, S.~Kumar and M.~Quartin,
``Constraints on cosmologically coupled black holes from gravitational 
wave observations and minimal formation mass,''
Mon. Not. Roy. Astron. Soc. \textbf{528}, no.2, 2377-2390 (2024)
doi:10.1093/mnras/stae143
[arXiv:2307.02474 [astro-ph.CO]].

\bibitem{Lacy:2023kbb}
M.~Lacy, A.~Engholm, D.~Farrah and K.~Ejercito,
``Constraints on Cosmological Coupling from the Accretion History of 
Supermassive Black Holes,''
Astrophys. J. Lett. \textbf{961}, no.2, L33 (2024)
doi:10.3847/2041-8213/ad1b5f
[arXiv:2312.12344 [astro-ph.CO]].




}

\bibitem{Boehm:2020jwd}
C.~Boehm, A.~Kobakhidze, C.~A.~J.~O'hare, Z.~S.~C.~Picker and 
M.~Sakellariadou,
``Eliminating the LIGO bounds on primordial black hole dark matter,''
JCAP \textbf{03}, 078 (2021)
doi:10.1088/1475-7516/2021/03/078
[arXiv:2008.10743 [astro-ph.CO]].

\bibitem{Harada:2021xze}
T.~Harada, H.~Maeda and T.~Sato,
``Thakurta metric does not describe a cosmological black hole,''
Phys. Lett. B \textbf{833}, 137332 (2022)
doi:10.1016/j.physletb.2022.137332
[arXiv:2106.06651 [gr-qc]].

\bibitem{Hutsi:2021nvs}
G.~H\"utsi, T.~Koivisto, M.~Raidal, V.~Vaskonen and H.~Veerm\"ae,
``Cosmological black holes are not described by the Thakurta metric: 
LIGO-Virgo bounds on PBHs remain unchanged,''
Eur. Phys. J. C \textbf{81}, no.11, 999 (2021)
doi:10.1140/epjc/s10052-021-09803-4
[arXiv:2105.09328 [astro-ph.CO]].

\bibitem{McVittie:1933zz}
G.~C.~McVittie,
``The mass-particle in an expanding universe,''
Mon. Not. Roy. Astron. Soc. \textbf{93}, 325-339 (1933)
doi:10.1093/mnras/93.5.325

\bibitem{Einstein:1945id}
A.~Einstein and E.~G.~Straus,
``The influence of the expansion of space on the gravitation fields 
surrounding the individual stars,''
Rev. Mod. Phys. \textbf{17}, 120-124 (1945)
doi:10.1103/RevModPhys.17.120

\bibitem{Faraoni:2015ula}
V.~Faraoni, {\em Cosmological and Black Hole Apparent Horizons}, 
Lect. Notes Phys. \textbf{907}  (Springer, New York, 2015),
ISBN 978-3-319-19239-0, 978-3-319-19240-6, 
doi:10.1007/978-3-319-19240-6

\bibitem{Croker:2021duf}
K.~S.~Croker, M.~J.~Zevin, D.~Farrah, K.~A.~Nishimura and G.~Tarle,
``Cosmologically Coupled Compact Objects: A Single-parameter Model for 
LIGO\textendash{}Virgo Mass and Redshift Distributions,''
Astrophys. J. Lett. \textbf{921}, no.2, L22 (2021)
doi:10.3847/2041-8213/ac2fad
[arXiv:2109.08146 [gr-qc]].


\bibitem{Croker:2019mup}
K.~S.~Croker and J.~L.~Weiner,
``Implications of Symmetry and Pressure in Friedmann Cosmology. I. 
Formalism,''
Astrophys. J. \textbf{882}, no.1, 19 (2019)
doi:10.3847/1538-4357/ab32da
[arXiv:2107.06643 [gr-qc]].

\bibitem{Croker:2019kje}
K.~Croker, K.~Nishimura and D.~Farrah,
``Implications of Symmetry and Pressure in Friedmann Cosmology. II. 
Stellar Remnant Black Hole Mass Function,''
doi:10.3847/1538-4357/ab5aff
[arXiv:1904.03781 [astro-ph.CO]].

\bibitem{Dahal:2023suw}
P.~K.~Dahal, F.~Simovic, I.~Soranidis and D.~R.~Terno,
``Black holes as spherically-symmetric horizon-bound objects,''
Phys. Rev. D \textbf{108} (2023) no.10, 104014
doi:10.1103/PhysRevD.108.104014
[arXiv:2303.15793 [gr-qc]].

\bibitem{Gaur:2023hmk}
R.~Gaur and M.~Visser,
``Black holes embedded in FLRW cosmologies,''
[arXiv:2308.07374 [gr-qc]].


\bibitem{Davidson:2012si}
A.~Davidson, S.~Rubin and Y.~Verbin,
``Can an evolving Universe host a static event horizon?,''
Phys. Rev. D \textbf{86}, 104061 (2012)
doi:10.1103/PhysRevD.86.104061
[arXiv:1210.0753 [gr-qc]].

{ 
\bibitem{Abreu:2010ru}
G.~Abreu and M.~Visser,
``Kodama time: Geometrically preferred foliations of spherically 
symmetric spacetimes,''
Phys. Rev. D \textbf{82}, 044027 (2010)
doi:10.1103/PhysRevD.82.044027
[arXiv:1004.1456 [gr-qc]].
}

\bibitem{Barrabes:1991ng}
C.~Barrabes and W.~Israel,
``Thin shells in general relativity and cosmology: The Lightlike limit,''
Phys. Rev. D \textbf{43}, 1129-1142 (1991)
doi:10.1103/PhysRevD.43.1129

\bibitem{Poisson:2002nv}
E.~Poisson,
``A Reformulation of the Barrabes-Israel null shell formalism,''
[arXiv:gr-qc/0207101 [gr-qc]].

\bibitem{FabbriNavarro-Salas} A. Fabbri and J. Navarro-Salas, {\em 
Modeling Black Hole Evaporation} (Imperial College Press/World Scientific, 
London/Singapore, 2005).

\bibitem{Hawking:1973uf} S.~W.~Hawking and G.~F.~R.~Ellis, {\em The Large 
Scale Structure of Space-Time} (Cambridge University Press, Cambridge, 
1973), ISBN 978-0-521-20016-5, 978-0-521-09906-6, 978-0-511-82630-6, 
978-0-521-09906-6 doi:10.1017/CBO9780511524646

\bibitem{Wald:1984rg} R. M.~Wald, {\em General Relativity} (Chicago 
University Press, Chicago, 1987), 
doi:10.7208/chicago/9780226870373.001.0001

\bibitem{Faraoni:2007gq}
V.~Faraoni,
``The Hawking temperature of expanding cosmological black holes,''
Phys. Rev. D \textbf{76}, 104042 (2007)
doi:10.1103/PhysRevD.76.104042
[arXiv:0710.2122 [gr-qc]].

\bibitem{Saida:2007ru}
H.~Saida, T.~Harada and H.~Maeda,
``Black hole evaporation in an expanding universe,''
Class. Quant. Grav. \textbf{24}, 4711-4732 (2007)
doi:10.1088/0264-9381/24/18/011
[arXiv:0705.4012 [gr-qc]].

{ 

\bibitem{Senovilla:1997zz}
J.~M.~M.~Senovilla and R.~Vera,
``Impossibility of the Cylindrically Symmetric Einstein-Straus Model,''
Phys. Rev. Lett. \textbf{78}, 2284-2287 (1997)
doi:10.1103/PhysRevLett.78.2284

\bibitem{Mars:1998sq}
M.~Mars,
``Axially symmetric Einstein-Straus models,''
Phys. Rev. D \textbf{57}, 3389-3400 (1998)
doi:10.1103/PhysRevD.57.3389
[arXiv:gr-qc/0202087 [gr-qc]].

\bibitem{Mars:2013ooa}
M.~Mars, F.~C.~Mena and R.~Vera,
``Review on exact and perturbative deformations of the Einstein-Straus 
model: uniqueness and rigidity results,''
Gen. Rel. Grav. \textbf{45}, 2143-2173 (2013)
doi:10.1007/s10714-013-1574-1
[arXiv:1307.4371 [gr-qc]].
}

\end{thebibliography}
\end{document}